\documentclass{PoS}

\title{The Short Baseline Neutrino Oscillation Program at Fermilab}

\ShortTitle{SBN Program at Fermilab}

\author{\speaker{Matthew Bass}\thanks{For the MicroBooNE and SBND Collaborations.}\\
        University of Oxford\\

        E-mail: \email{Matthew.Bass@physics.ox.ac.uk}}

\abstract{The Short-Baseline Neutrino (SBN) Program is a short-baseline neutrino oscillation experiment in the Booster Neutrino Beam-line (BNB) at Fermilab. It consists of three Liquid Argon Time Projection Chambers (LArTPCs) from the Short-Baseline Near Detector (SBND), Micro Booster Neutrino Experiment (MicroBooNE), and Imaging Cosmic And Rare Underground Signals (ICARUS) experiments. The SBN Program will definitively search for short-baseline neutrino oscillations in the 1 eV mass range, make precision neutrino-argon interaction measurements, and further develop the LArTPC technology. The physics program and current status of the program, and its constituent experiments, are presented.}

\FullConference{38th International Conference on High Energy Physics\\
                3-10 August 2016\\
                Chicago, USA}

\begin{document}

\section{Introduction}

\begin{figure}[ht!]
\begin{center}
\includegraphics[width=0.85\linewidth]{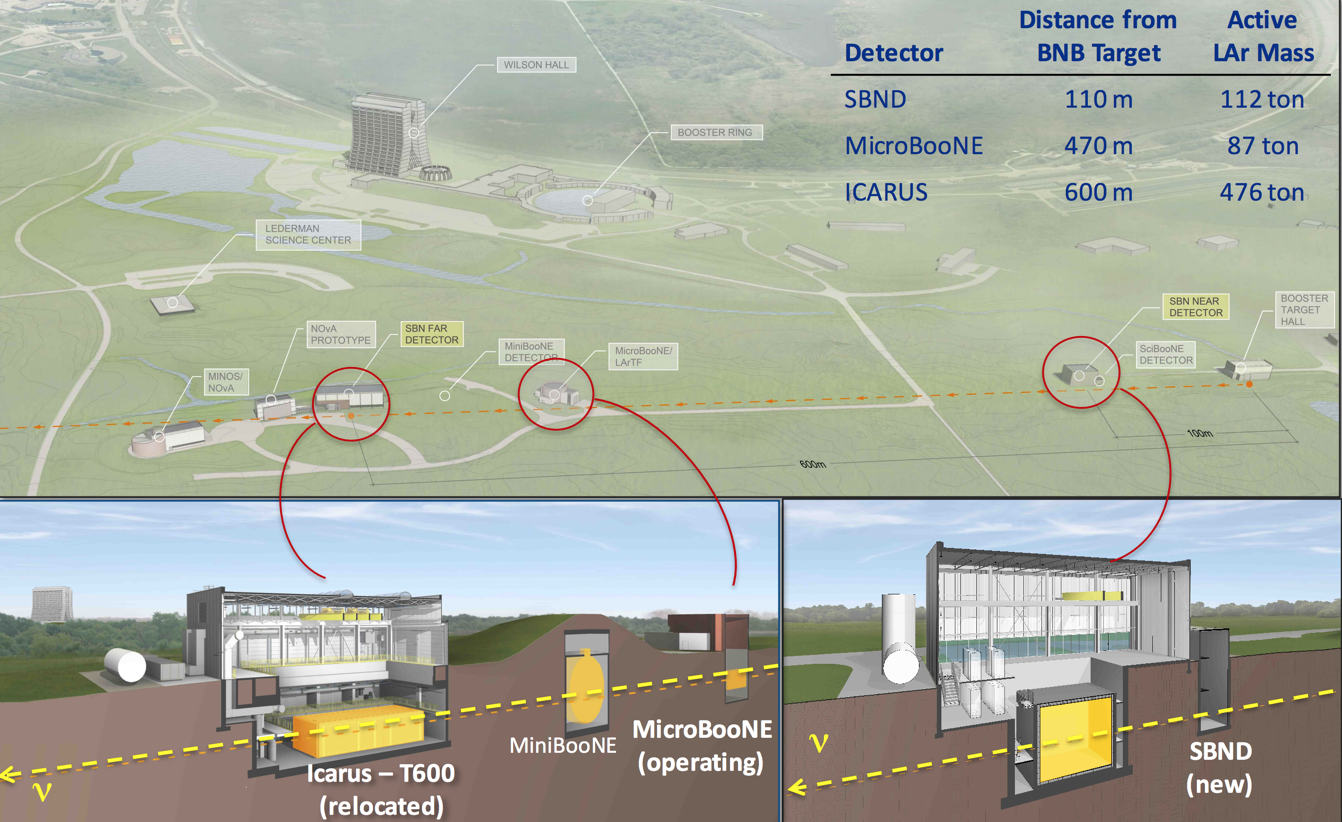}
\end{center}
\caption{Layout and active detector masses of the three LArTPCs of the SBN Program and their orientation with respect to the BNB target.} \label{fig:layout}
\end{figure}

The Short-Baseline Neutrino (SBN) Program is an extensive experimental program to explore neutrino properties and detector technology development in a neutrino beam-line. The SBN program consists of three Liquid Argon Time Projection Chambers (LArTPCs), placed at varying distances from the neutrino source, running in the Booster Neutrino Beam-line (BNB) at Fermilab near Chicago, IL. It is a collection of three experimental collaborations and consists of the SBND, MicroBooNE, and ICARUS detectors configured as depicted in Figure \ref{fig:layout}. The goals of the program are to followup on the MiniBooNE low energy excess, definitively explore the phase space of short-baseline neutrino oscillations, make precision measurement of neutrino-argon interactions, and further develop the LArTPC technology.

\section{SBN Program Goals}
\subsection{Sterile $\mathbf{\nu}$ Oscillations}

The primary physics goal of the SBN Program is to follow up on hints of new physics seen in past experiments\cite{lsnd}\cite{mb}. These experiments reported results that could be indicative of one or more new sterile neutrino states in the 1 eV mass range. These new sterile neutrino states would cause oscillations resulting in electron neutrino appearance at short-baselines ($< 1$ km) in pion-decay in flight based neutrino beams, such as the BNB. 

\begin{figure}[h!]
\begin{center}
\includegraphics[width=0.48\linewidth]{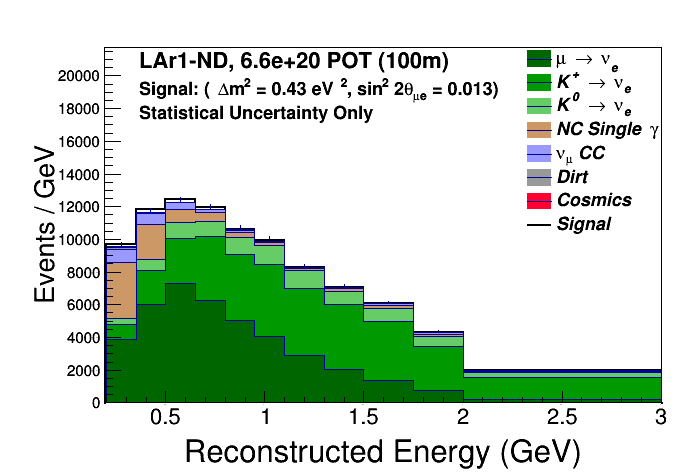}
\includegraphics[width=0.48\linewidth]{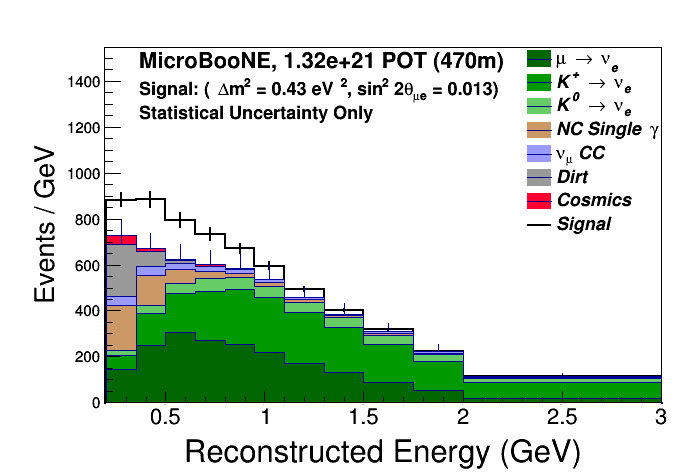}
\\
\includegraphics[width=0.48\linewidth]{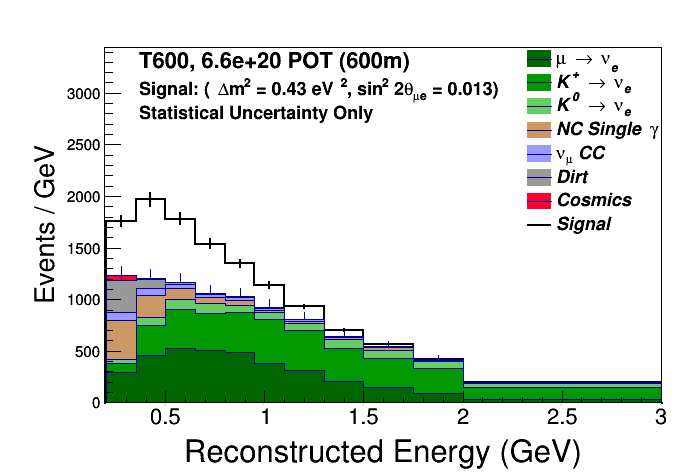}
\end{center}
\caption{Simulated event spectra for charge-current electron neutrino events at the three SBN Program detectors with backgrounds in stacked histograms\cite{Antonello:2015lea}. An injected signal, in black, uses the values of $\Delta \mathrm{m}^2$ and $\sin^2 2\theta_{\mu \mathrm{e}}$ from a global fit\cite{Kopp:2013vaa} and demonstrates the expected signal enhancement with increasing baseline.} \label{fig:spectra}
\end{figure}

The size and placement of the three LArTPCs in the SBN Program allow for characterization of the beam before oscillations in SBND and simultaneous measurement of $\nu_\mathrm{e}$ appearance and $\nu_\mu$ disappearance in the MicroBooNE and ICARUS detectors. Event spectra from a detailed simulation of each detector, with an injected signal at global best fit values, are shown in Figure \ref{fig:spectra}. 

\begin{figure}[ht!]
\begin{center}
\includegraphics[width=0.48\linewidth]{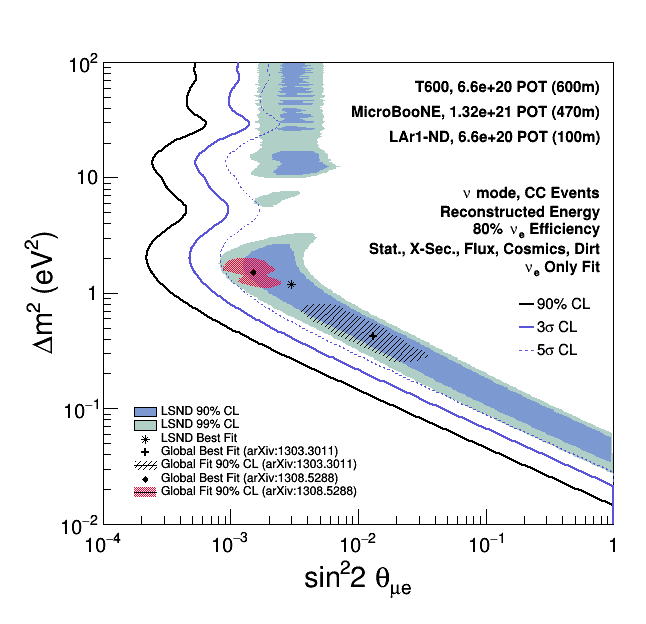}
\end{center}
\caption{SBN Program sensitivity to short-baseline neutrino oscillations in $\Delta m^2$ vs $\sin^2 2\theta_{\mu \mathrm{e}}$. The LSND\cite{lsnd} and two sets of global fit\cite{Kopp:2013vaa}\cite{Giunti:2013aea} confidence intervals and best fit points are plotted for reference.  }
\label{fig:sense}
\end{figure}

The expected sensitivity for a combined, three-detector analysis as a function of $\Delta \mathrm{m}^2$ and $\sin^2 \theta_{\mu \mathrm{e}}$ is shown in Figure \ref{fig:sense} overlaid with the allowed regions and best fits from LSND and multiple global fits\cite{Kopp:2013vaa}\cite{Giunti:2013aea}. The signal region for the LSND and global best fits are all projected to be excluded at least at the $5\sigma$ level using the full data set from all three detectors.

Recent global fits\cite{Collin:2016aqd} using the 3+1 sterile neutrino model have placed further constraints on the allowed range of values for the oscillation parameters governing the $\nu_\mu \to \nu_e$ oscillation over short baselines, excluding the $\Delta \mathrm{m}^2 = 1$~eV$^2$ solution at 90\% CL.
The configuration of the SBN Program will enable measurement of oscillations at the $5\sigma$ level for this solution. 
Furthermore, the SBN Program will explore these short-baseline oscillations in a single experiment, obviating the need for a global fit over multiple experiments to reach sensitivity in the $\Delta \mathrm{m}^2 \sim 1$~eV$^2$ region. This will give critical information to future long-baseline experiments where sensitivity to CP violation can be reduced if 3+1 or 3+2 sterile neutrino models are used to fit the data\cite{Gandhi:2015xza}. 

\subsection{MiniBooNE Low Energy Excess}
The MiniBooNE experiment ran also in the BNB and has published results\cite{mblee} of excesses of electron neutrino like events with $3.4\sigma$ ($2.8\sigma$) significance in neutrino (anti-neutrino) running modes.
The neutrino flux from the BNB has been studied extensively and its electron-neutrino contamination is well known \cite{mbflux}. The source and nature of this excess is still an open question.

Because the MiniBooNE experiment used the Cerenkov technology to distinguish between muon and electron induced signals, it was not able to easily distinguish between electron and photon induced interactions. The LArTPC technology allows the topology and calorimetric properties of an electron interaction to be distinguished from that of a photon. This is achieved by either observing a gap between a neutrino interaction vertex and a shower or by observing calorimetric differences in the beginning of a shower between a $\gamma \to e^+e^-$ pair production and a single ionizing electron. This electron-gamma separation technique was used in studies by the ArgoNeuT experiment\cite{Acciarri:2016sli} to make the first observation of low-energy electron neutrinos in a neutrino beam-line.  A primary goal of the MicroBooNE experiment is to follow up on this low energy excess. Using this technique, it will provide a definitive answer about the nature of these events.

\subsection{Neutrino-Argon Interactions}

Future long-baseline neutrino experiments\cite{Acciarri:2015uup} will require significant improvements to neutrino interaction modeling in order to make precision measurements of neutrino oscillations. The SBN Program will make the highest precision cross section measurements of $\nu_e$-Ar and $\nu_\mu$-Ar scattering in the few hundred MeV to few GeV range as each of the detectors see neutrino interactions coming from the on-axis BNB and off-axis NuMI beams.  

\begin{figure}[ht!]
\begin{center}
\includegraphics[width=0.48\linewidth]{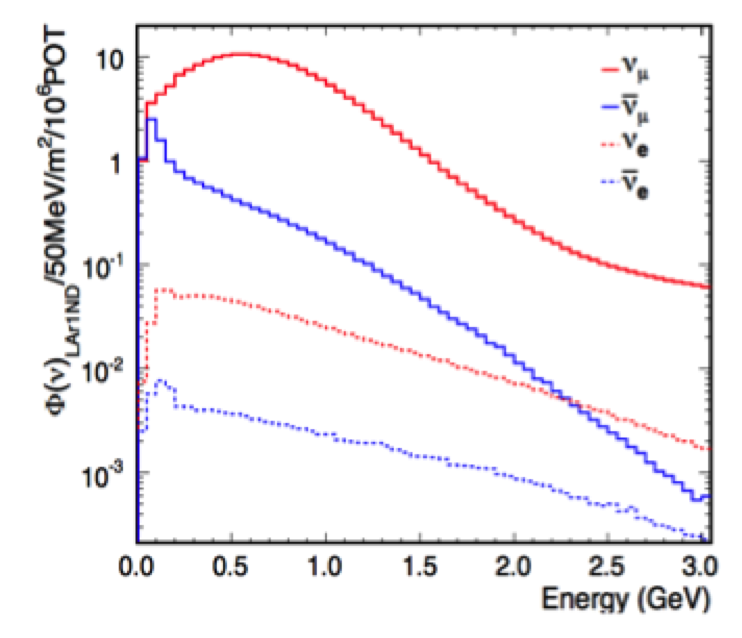}
\includegraphics[width=0.48\linewidth]{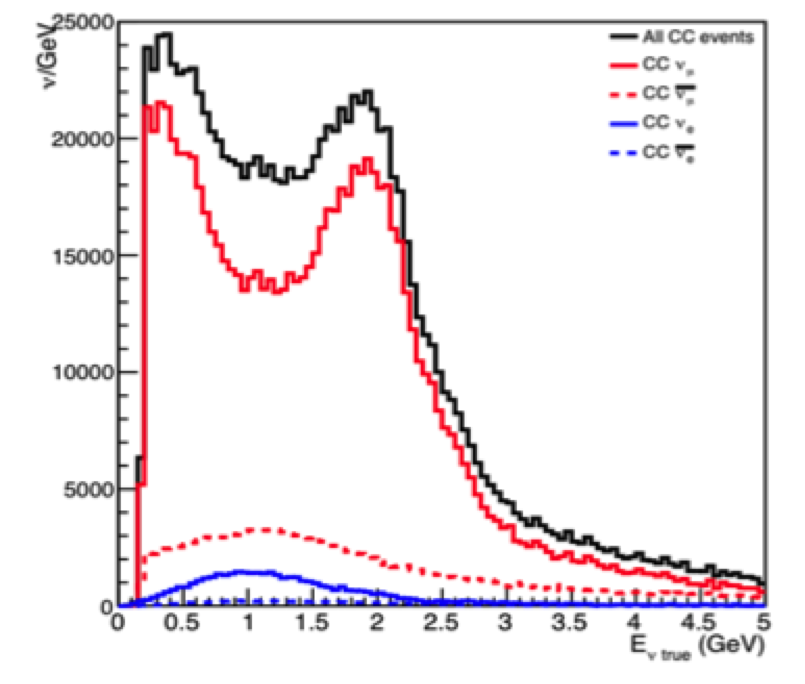}
\end{center}
\caption{Simulated event spectra (flux times cross section) for neutrino interactions at SBND (left) from the BNB and ICARUS detector (right) from the NuMI neutrino beam-line\cite{Antonello:2015lea}.} \label{fig:flux}
\end{figure}

In the near term, MicroBooNE will measure $\nu_\mu$-Ar ($\nu_e$-Ar) cross sections with hundreds of thousands (thousands) of interactions expected over the course of the three year initial run. SBND will improve on these statistics dramatically, with $1.5\times 10^6$ ($1.2\times 10^4$) $\nu_\mu$-Ar ($\nu_e$-Ar) interactions expected over the course of the experiment, from the BNB neutrino flux, shown in Figure \ref{fig:flux}. Interesting exclusive channel measurements, such as coherent scattering, strange production, and neutrino-electron scattering, will also be possible due to the high event rate at the near detector. The MicroBooNE and ICARUS detectors each also see interactions from the Neutrinos in Main Injector (NuMI) neutrino beam-line at off-axis angles. ICARUS will see greater than 10,000 $\nu_e$-Ar interactions per year in the 1 GeV region from the NuMI beam, as depicted in Figure \ref{fig:flux} where the flux is significantly peaked due to off-axis effects.

\subsection{LArTPC Detector Development}
The LArTPC detector technology was chosen because of its excellent spatial resolution and calorimetry which allows for efficient separation of electron and gamma signals based on topology and energy deposition.
LAr is also advantageous as a detector medium because of the large neutrino-argon cross section, its transparency to its own scintillation light, and its relative abundance and low cost.

The LArTPC technology will be used for the Deep Underground Neutrino Experiment\cite{Acciarri:2015uup} in which large, kilotonne-scale LArTPCs will be needed to achieve sufficient sensitivity to CP violation.
The SBND design includes features that are similar to those planned for the DUNE far detector and so will be a critical proof of concept of these designs.
Scaling up to these large detector masses will benefit from the developments in, for example, LAr purification, modular detector anode components, and scintillation-light detection systems used in the SBN program.

In addition, simulation, reconstruction, and event selection algorithms are being developed in a shared software platform for use by all LArTPC based experiments\cite{Church:2013hea}. These reconstruction and event selection algorithms are critical for the success of both the SBN Program and for DUNE.

\section{SBN Program Status}
The MicroBooNE experiment began taking data in August 2015, has recorded $3.4\times10^{20}$ Protons-On-Target (POT) as of August 2016, and will continue taking data with an expected total exposure of $6.6\times10^{20}$ POT. A cosmic ray tagger system has been installed over the summer of 2016 to better enable cosmic ray background removal at the analysis stage. The MicroBooNE collaboration has presented its first analyses\cite{publicnotes} and is in the process of preparing multiple publications\cite{Acciarri:2016ryt}.

The civil construction of the facilities for the SBND experiment is underway and components of the detector are under construction. A cosmic ray tracker system is being built and tested. TPC assembly will begin at Fermilab in 2017 and it will be installed in the cryostat in 2018. Commissioning is planned for 2018 and operations will begin in 2019.

The civil construction for the facilities for the ICARUS detector is also underway and will be completed by the end of 2016. 
The ICARUS TPCs are undergoing refurbishment and being upgraded at CERN. This includes ensuring cathode planarity and refurbishing the cryogenic system. 
At the same time the optical system is being enlarged and the detector electronics upgraded.
These TPCs will be delivered to Fermilab in early 2017 and commissioning will begin later that year with operations beginning in 2018.

\section{Conclusions}

The SBN Program, consisting of the SBND, MicroBooNE, and ICARUS experiments, will definitively explore the allowed phase space for short-baseline neutrino oscillations, follow up on the MiniBooNE low-energy excess, and make many precision neutrino-argon cross-section measurement in the BNB and NuMI neutrino beam-lines at Fermilab. With its three LArTPCs, it will also develop the LArTPC detector technology necessary for the DUNE experiment. The MicroBooNE experiment is taking data currently, while SBND and ICARUS are under construction and will begin taking data in 2018.

\end{document}